**Tri-modality Cavitation Mapping in Shock Wave Lithotripsy**

Mucong Li,[1] Georgy Sankin,[2] Tri Vu,[1] Junjie Yao,[1,a] and Pei Zhong[2,b]

[1] *Department of Biomedical Engineering, Duke University, Durham, NC 27708, USA*

[2] *Department of Mechanical Engineering and Materials Science, Duke University, Durham, NC 27708, USA*

Shock wave lithotripsy (SWL) has been widely used for non-invasive treatment of kidney stones. Cavitation plays an important role in stone fragmentation, yet may also contribute to renal injury during SWL. It is therefore crucial to determine the spatiotemporal distributions of cavitation activities to maximize stone fragmentation while minimizing tissue injury. Traditional cavitation detection methods include high-speed optical imaging, active cavitation mapping (ACM), and passive cavitation mapping (PCM). While each of the three methods provides unique information about the dynamics of the bubbles, PCM has most practical applications in biological tissues. To image the dynamics of cavitation bubble collapse, we previously developed a sliding-window PCM (SW-PCM) method to identify each bubble collapse with high temporal and spatial resolution. To further validate and optimize the SW-PCM method, in this work, we have developed tri-modality cavitation imaging that includes 3D high-speed optical imaging, ACM, and PCM seamlessly integrated in a single system. Using the tri-modality system, we imaged and analyzed laser-induced single cavitation bubbles in both free and constricted space and shockwave-induced cavitation clusters. Collectively, our results have demonstrated the high reliability and spatial-temporal accuracy of the SW-PCM approach, which paves the way for the future *in vivo* applications on large animals and humans in SWL.

**Key words** – Shockwave lithotripsy, cavitation, passive cavitation detection

[a] junjie.yao@duke.edu
[b] pei.zhong@duke.edu

I. **INTRODUCTION**

Kidney stone disease is a major health problem (Mortality & Causes of Death, 2015; Pearle, Calhoun, Curhan, & Urologic Diseases of America, 2005; Scales et al., 2007; Stamatelou, Francis, Jones, Nyberg, & Curhan, 2003; Taylor, Stampfer, & Curhan, 2005). In the U.S., approximately 9% of the population will experience an episode of kidney stone formation in their lifetime. Overall, a significant economic burden is associated with kidney stones, with healthcare costs exceeding $5 billion annually (Hyams & Matlaga, 2014). Shock wave lithotripsy (SWL) remains one of the first-line therapies for the management of most kidney stones (Assimos et al., 2016a, 2016b; Pearle, Lingeman, et al., 2005; Preminger et al., 2007) largely due to its noninvasive nature, ease of use, and high reimbursement rate by medical plans (Lotan, Cadeddu, Roehrborn, & Stage, 2004; Pearle, Lingeman, et al., 2005). A shock wave lithotripter generates focused high-energy shockwave pulses, which are transmitted through patient's skin surface and interposed tissues to disintegrate stones located in the kidney or the upper urinary track. The comminution of kidney stones involves two main contributing factors. The first one is produced by the interaction of the leading compressive pressure of the shockwave pulse with the target stone, leading to the generation of tensile and shear stresses and resultant fracture of the stone materials into smaller pieces (Cleveland & Sapozhnikov, 2005; Eisenmenger, 2001; Lingeman, Kim, Kuo, McAteer, & Evan, 2003; Xi & Zhong, 2001). The second factor is related to the trailing tensile pressure of the shockwave pulse, which causes the formation of cavitation bubbles around the stone surfaces. The violent collapses of the cavitation bubbles generate surface pitting (Church, 1989; Coleman, Saunders, Crum, & Dyson, 1987; Sass et al., 1991) and, potentially through synergistic interaction with stress waves, produce fine fragments to facilitate spontaneous discharge (Coleman et al., 1987; Zhu, Cocks, Preminger, & Zhong, 2002). While cavitation plays an important role in stone comminution, it has also been identified as a major cause of renal injury (Matlaga et al., 2008; Zhong, Zhou, & Zhu, 2001). Vascular injuries during SWL could

be produced by the rapid intraluminal bubble expansion in capillaries or small blood vessels or by the microjets formed due to asymmetric bubble collapse (Pishchalnikov et al., 2003; Treglia & Moscoloni, 1999; Zhong et al., 2001). Acute renal hemorrhage and damage to the surrounding renal tubules have been widely reported in both animal and human studies, including severe and permanent renal damage or even death in extreme cases (Stoller, Litt, & Salazar, 1989; Toro & Kardos, 2008; Treglia & Moscoloni, 1999). It is therefore crucial to monitor the cavitation dynamics in SWL to both optimize the stone comminution efficiency and minimize the risk of tissue injury.

Both optical and acoustic methods have been used for cavitation detection. High-speed photography can be applied to capture bubble distribution *in vitro*, but it is only applicable to optically transparent media (Sankin, Simmons, Zhu, & Zhong, 2005). Acoustic methods include active cavitation mapping (ACM) and passive cavitation mapping (PCM) (Cleveland, Sapozhnikov, Bailey, & Crum, 2000; Gateau, Aubry, Pernot, Fink, & Tanter, 2011; Madanshetty, Roy, & Apfel, 1991). For *in vivo* scenarios, in ACM the backscattered ultrasound signals from the cavitation bubble are usually obliterated by the background tissue signals. Although algorithms such as decorrelation and singular value decomposition (SVD) filtering can be applied to separate bubble signals from the background signals, the motion artifacts still significantly degrade the image contrast (Bader, Vlaisavljevich, & Maxwell, 2019; Desailly et al., 2017). In addition, ACM is not sensitive to bubble collapse, which may correlate with renal hemorrhage. Therefore, PCM is often the preferred method for cavitation detection *in vivo* with cleaner background, superior contrast and less artifacts.

The essential task in PCM is to reconstruct the cavitation distribution from the raw signals emitted by individual bubble collapses. Different PCM methods have been used in a variety of medical applications such as high intensity focused ultrasound (HIFU), histotripsy, and acoustic angiography (T. Li, Khokhlova, Sapozhnikov, O'Donnell, & Hwang, 2014; Rojas et al., 2019; Vlaisavljevich et al., 2014). For example, time exposure acoustics (TEA) projects signals back into the image space and averages over time to enhance the acoustic sources at fixed positions (Norton & Won, 2000). Temporal or frequency domain passive projection shifts the channel data based on the travel time of the acoustic signals to reconstruct cavitation distribution, assuming that all bubbles collapse at the same time (Acconcia, Jones, Goertz, O'Reilly, & Hynynen, 2017; Haworth, Bader, Rich, Holland, & Mast, 2017; Xu et al., 2019). These two methods perform well for HIFU in which the bubbles generated are typically small and tightly confined within a small region, and they collapse almost instantaneously after the cessation of acoustic excitation. In addition, the bubble cloud distribution (rather than the individual bubbles) are more relevant in HIFU applications. However, the shockwave lithotripter focus is typically much larger than that of the HIFU beam. The bubbles distribute sparsely in space and collapse asynchronously at different times (spanning over hundreds of microseconds following shockwave excitation). Therefore, the aforementioned PCM methods cannot be applied to characterize SWL-induced cavitation.

In this study, we have developed a sliding-window PCM (SW-PCM) algorithm modified from the classic delay-and-sum method widely used in ultrasound and photoacoustic imaging. The temporal randomness of shockwave induced bubble collapses is taken into consideration during the cavitation reconstruction. Each individual bubble collapse can be precisely reconstructed temporally and spatially. We performed cavitation mapping of both laser-induced single bubbles and lithotripter shockwave induced bubble clusters. More importantly, we have constructed an integrated tri-modality system, and for the first time, the cavitation's temporal and spatial distributions by PCM were validated concurrently using a 3D high-speed camera and ACM, respectively. The SW-PCM results have an accuracy of more than 90% for single bubbles and 83% for bubble clusters. These tri-modality results collectively demonstrate the reliability of our SW-PCM method.

## II. METHODS

### A. The tri-modality cavitation mapping system

The schematic of the tri-modality cavitation mapping system is shown in **Fig. 1**(a), with the single-bubble cavitation generated by a pulsed laser (Sankin et al., 2005). The 1064 nm light beam from a pulsed laser (Surelite lasers, Continuum, CA; Pulse repetition rate: 10 Hz) was first expanded and then focused by a convex lens with a large numerical aperture (NA = 1) into a water tank, with an optical focal spot size of ~0.81 µm. A single bubble was produced at the optical focus when the laser pulse fluence exceeded the optical breakdown threshold of water (Visscher, Lajoinie, Blazejewski, Veldhuis, & Versluis, 2019). A 128-element linear-array ultrasonic transducer (L7-4, Philips; Central frequency: 5.208 MHz) was placed 30 mm above the optical focus of the laser beam to detect the cavitation signals. The imaging plane of the transducer was aligned with the laser beam focus. The cavitation signals were acquired by a programmable ultrasound scanner (Vantage 128, Verasonics), which was synchronized with the laser pulse firing. Both ACM and PCM were performed using the ultrasonic transducer., in which a 0.1 µs ultrasound pulse was transmitted to the optical focus region followed by 800 µs of signal recording with a sampling rate of 20.8 MHz. Both the reflected ultrasound signals by the bubble (ACM) and the bubble collapse signals (PCM) were received and reconstructed.

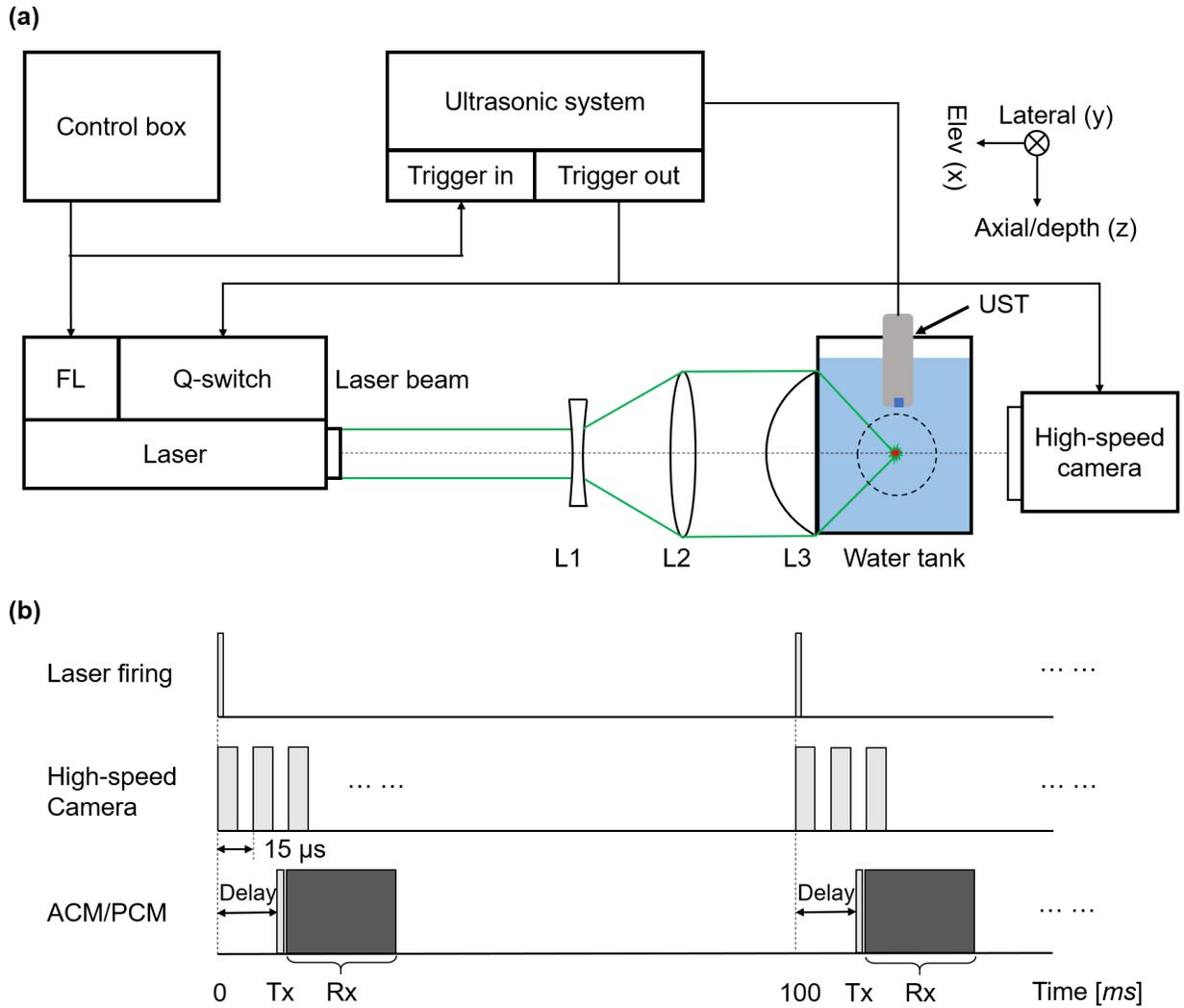

**Fig. 1**. Tri-modality cavitation mapping of laser-induced single bubbles. (a) System schematics. FL: flashlamp; L1, L2, L3: focusing lens. The displayed position of high-speed camera is only for drawing convenience and its actual position is shown by the dashed circle. (b) Time sequence of synchronized laser firing, high-speed camera recording, and ACM/PCM detection. UST: ultrasound transducer; Tx: ultrasound pulse transmission; Rx: ultrasound signal receiving; Elev: elevational.

The bubble dynamics were also captured simultaneously by a color high-speed camera (Phantom 7.3, Vision Research), which recorded a total of 55 frames after each laser pulse with a frame rate of 66 kHz. The camera's focal plane was co-aligned with the laser focal spot. An LED illuminator provided the light source for the camera. The time sequence of the laser firing, camera recording, and acoustic signal detection is shown in **Fig. 1**(b). The camera recording started 12 μs after the laser firing. The ultrasound transmission started right after the laser firing when imaging bubbles in free water space, and 60 μs after when imaging bubbles in constraint space. The PCM signals typically lagged the ACM signals by >200 μs, long enough to separate the two types of signals.

### B. Reconstruction of the ACM and PCM images

A bandpass filter (cutoff frequencies: 3.5 MHz and 6.5 MHz) was first applied to the raw channel data before the image reconstruction. In ACM, the reflected ultrasound signals by the bubbles were reconstructed using the conventional delay-and-sum method, with the ultrasound transmission moment designated as the time origin. For each bubble event, the ACM reconstructed image size was 40 mm by 40 mm with a pixel size of 67 μm. In PCM, the bubble bursts were considered as sparsely distributed point acoustic sources, and the bubble collapses were reconstructed by using our previously developed sliding-window reconstruction method (M. Li et al., 2020). Briefly, the reconstruction window, with a fixed window size, slides along the time axis to search for the correct time origin of each individual bubble collapse. The search stopped when the reconstructed signals maximally converged in space and achieved the highest acoustic energy density. At each reconstruction instance, *i.e.*, the sliding window position, the signal amplitude in the reconstructed PCM image can be written as (Gyongy & Coussios, 2010):

$$H(\vec{r_s}, t) = \sum_{n=1}^{128} \frac{a(\vec{r_n} - \vec{r_s})}{|\vec{r_n} - \vec{r_s}|} * p_n(t + \frac{|\vec{r_n} - \vec{r_s}|}{c}),$$

where $H(\vec{r_s})$ is the reconstructed signal amplitude at location $\vec{r_s}$ inside the imaging plane, *a* denotes the ultrasound probe's angular detection sensitivity, *n* is the index of the ultrasound probe element, $\vec{r_n}$ is the location of the *nth* probe element, *p* is the signal recorded by the *nth* element, *t* is the sliding window position, *i.e.*, delay after the laser pulse firing, and *c* is the speed of sound (1500 m/s in water). At each reconstruction instance, the PCM reconstructed image size was 40 mm by 40 mm with a pixel size of 65 μm. The sliding step size of the reconstruction window was 0.2 μs, which corresponded to a spatial resolution of 300 μm. An illustration of this sliding-window method is shown in **Supplementary Video 1**.

The whole searching process in PCM can be projected onto a single imaging plane (lateral-axial plane of ultrasound transducer), as shown in **Supplementary Fig. 1**(a). By sliding the reconstruction window, the PCM signals from individual bubbles converged at different times. Because the bubble bursts were considered as individual point sources in both space and time, the bubble collapse time was identified when the reconstructed cavitation signals were maximally converged. For each reconstruction instance, the energy density of the reconstructed signals was computed to quantify the level of convergence. The energy density was defined as the averaged intensity of the top ten pixels within the converging zone (**Supplementary Fig. 1**(b)).

**C. Tri-modality cavitation mapping of laser-induced single bubbles in free water space**

The accuracy of PCM-detected bubble collapse time and location was first validated by laser-induced single bubbles. To validate the collapse time, PCM results were compared with the high-speed camera recordings. Bubbles in each camera frame were extracted by applying a Matlab boundary detection algorithm and then projected as a spatial-temporal plot to show the dynamics. The true bubble collapse time captured by the camera was usually between the two frames with the minimal bubble sizes. The energy density of the PCM results was computed and compared with the corresponding camera recording. Four representative cavitation events were presented to demonstrate the system's repeatability.

To validate the collapse location, the PCM results were compared with the ACM results. The spatial distribution of the PCM- and ACM-detected bubbles were analyzed, and the discrepancies were quantified. The reconstructed bubble locations by the two modalities were combined into a 3D space-time matrix and rendered by using Volview (Kitware, NY). A total of 30 cavitation events were analyzed to demonstrate the system's repeatability.

**D. Tri-modality cavitation imaging of laser-induced bubbles in a constrained space**

To mimic cavitation in blood vessels, a constrained environment, the laser beam was focused into the cross-sectional center of a transparent plastic tube through a small opening on the tube wall (S3TM-E3603, Saint-Gobain Tygon; ID, 2 mm; OD, 3 mm). Both high-speed camera imaging and ACM were acquired to validate the PCM results. To enhance the ultrasound reflection by the bubbles in ACM, the ultrasound transmission was performed 60 μs after the laser firing, which allowed the bubbles to expand. To improve the image contrast of the bubbles, ACM without laser excitation was first performed to obtain the background signals from the tube wall, which were then subtracted from the ACM images with laser excitation. A total of 40 cavitation events were statistically analyzed. Other settings and data processing methods were the same as in the free water space experiment.

**E. Tri-modality cavitation imaging of shockwave-induced bubble clusters**

After validating the tri-modality cavitation mapping system on laser-induced single bubbles, we further applied the system to study the lithotripter shockwave induced cavitation clusters. Shockwave pulses were generated by a focused shockwave generator (Piezoson 100, Richard Wolf, Germany), which was mounted at the bottom of the water tank. The shockwave generator transmitted microsecond shockwave pulses with a pulse repetition rate of 1 Hz and a focal distance of 65 mm. The resultant cavitation bubble clusters distributed sparsely in a volumetric space around the focal zone, which was approximately 1 cm laterally by 2 cm axially.

Again, we used the high-speed camera and the linear-array ultrasound probe to capture the shockwave-induced bubble clusters. With a relatively large depth of focus, the high-speed camera can project all the bubbles throughout the whole shockwave focal zone into a single 2D image. However, the ACM and PCM can only detect bubbles located within the ultrasonic transducer's imaging plane, i.e., a thin 2D slice cutting through the shockwave focal zone. In other words, the ultrasonic transducer would only detect a portion of the bubbles captured by the high-speed camera. Therefore, it is challenging to directly correlate cavitation bubbles captured by the high-speed camera and the ultrasonic transducer. To address this issue, we implemented a 3D camera system (F. Gao et al., 2017). As shown in **Fig. 2**(a), we constructed two orthogonal white-light illumination beams that first passed through the shockwave focal zone, and then transmitted through respectively a red filter and a blue filter. The filtered light beams were combined by a dichroic mirror before captured by a color CCD camera. The red and blue channels of the camera images were recorded separately. The red channel image provided the x-z projection of the bubbles, and the blue channel provided the y-z projection. A 3D reconstruction of the bubble cluster was extracted from the two projections. We aligned the imaging plane of ultrasonic transducer through the shockwave focus, with an angle of 45° relative to both illumination paths. The lateral direction of the ultrasonic transducer corresponds to the z direction of camera captured images (**Fig. 2**(b)). From the 3D reconstruction of the camera-captured bubbles, we extracted the group of bubbles within the ultrasonic transducer's imaging plane to validate the ACM/PCM images. The time sequence of shockwave transmission, camera recording, and ACM/PCM detection is shown in **Fig. 2**(c). The camera imaging and shockwave transmission were synchronized. A total of 55 camera frames were captured for each shockwave pulse, with a frame rate of 50 kHz. To acquire the ACM/PCM images, we applied a delay time of ~50 μs after the shockwave transmission, allowing the shockwaves to completely pass through the ACM/PCM imaging plane and avoiding the acoustic signal interference. The delay also allowed the bubbles to grow substantially, providing stronger ACM signals.

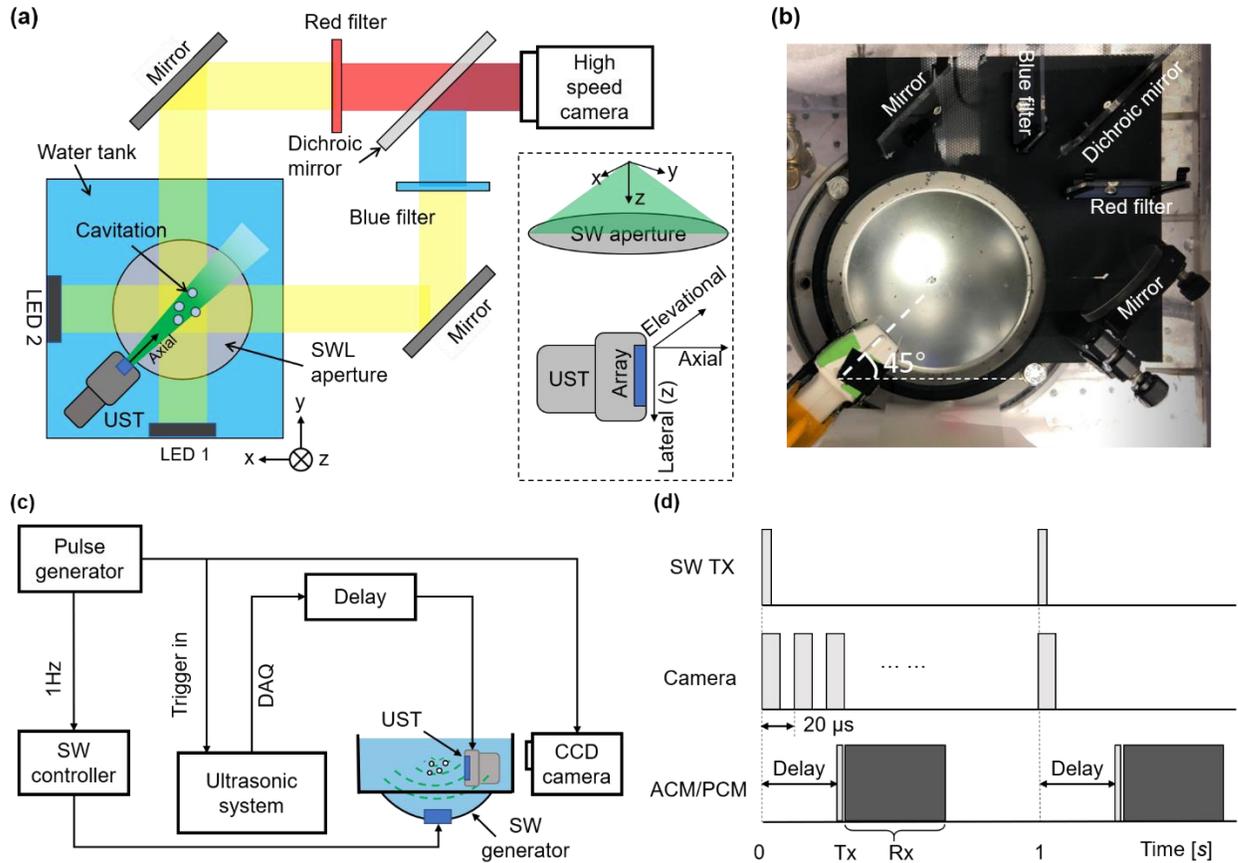

**Fig. 2.** Tri-modality cavitation mapping system of lithotripter shockwave-induced bubble cluster. (a) Schematics of the experimental setup, showing the two orthogonal illumination beams and the 45-degree ultrasonic transducer. UST: ultrasonic transducer. The relative position of the lithotripter aperture and the ultrasonic transducer is shown in the dashed rectangle. (b) Photograph of experimental setup. (c) Time sequence of shockwave transmission, camera recording, and ACM/PCM detection. SW TX: shockwave transmission; Tx: ultrasound transmission; Rx: ultrasound/cavitation signal receiving.

### F. Data analysis of shockwave-induced bubble clusters

The bubble-extraction algorithm used in laser-induced single bubble detection was applied here to quantify the bubble size and location in the 3D camera images. Distinctively separated and relatively large bubbles from the camera images were chosen for further analysis and comparison with the ACM/PCM results. To determine the group of bubbles within the image plane of the ultrasonic transducer, we computed the acoustic detection sensitivity in the imaging plane, using the ultrasound simulation toolbox Field II (Jensen, 1991). The camera-captured bubbles in the ultrasound imaging plane were then compared with the ACM and PCM images. Same as the laser-induced single bubble experiment, the ACM and PCM images were assembled in 3D space-time rendering to better visualize the cavitation distribution and dynamics.

### III.    RESULTS AND DISCUSSION

**A. Tri-modality cavitation mapping of laser-induced single bubbles in free water space**

**Figure 3**(a) shows representative frames of single bubble dynamics captured by the high-speed camera with a frame interval of 15 µs. The first frame was taken 12 µs after the laser firing due to the camera exposure. The bubble expanded to ~2 mm in diameter at 132 µs, and then quickly shrank and eventually collapsed between 207 µs and 222 µs. The primary bubble collapse produced two smaller daughter bubbles. **Figure 3**(b) shows the dynamics of a total of four representative single bubble events captured by the camera from 100 µs to 330 µs, from which we were able to identify the bubble collapse time window as marked by the red lines between two consecutive frames. Meanwhile, we quantified the acoustic energy density profile as a function of time from the SW-PCM reconstruction images, where the time point of the peak energy density reflects the collapse instant of the bubble, as indicated by the dotted black lines in **Fig. 3**(c). The bubble collapse times detected by the camera and PCM were consistent for all the four bubble events. It is worth noting that the PCM detection of the bubble collapse time has a temporal resolution of 0.2 µs, which is 75 times better than that of high-speed camera (~15 µs).

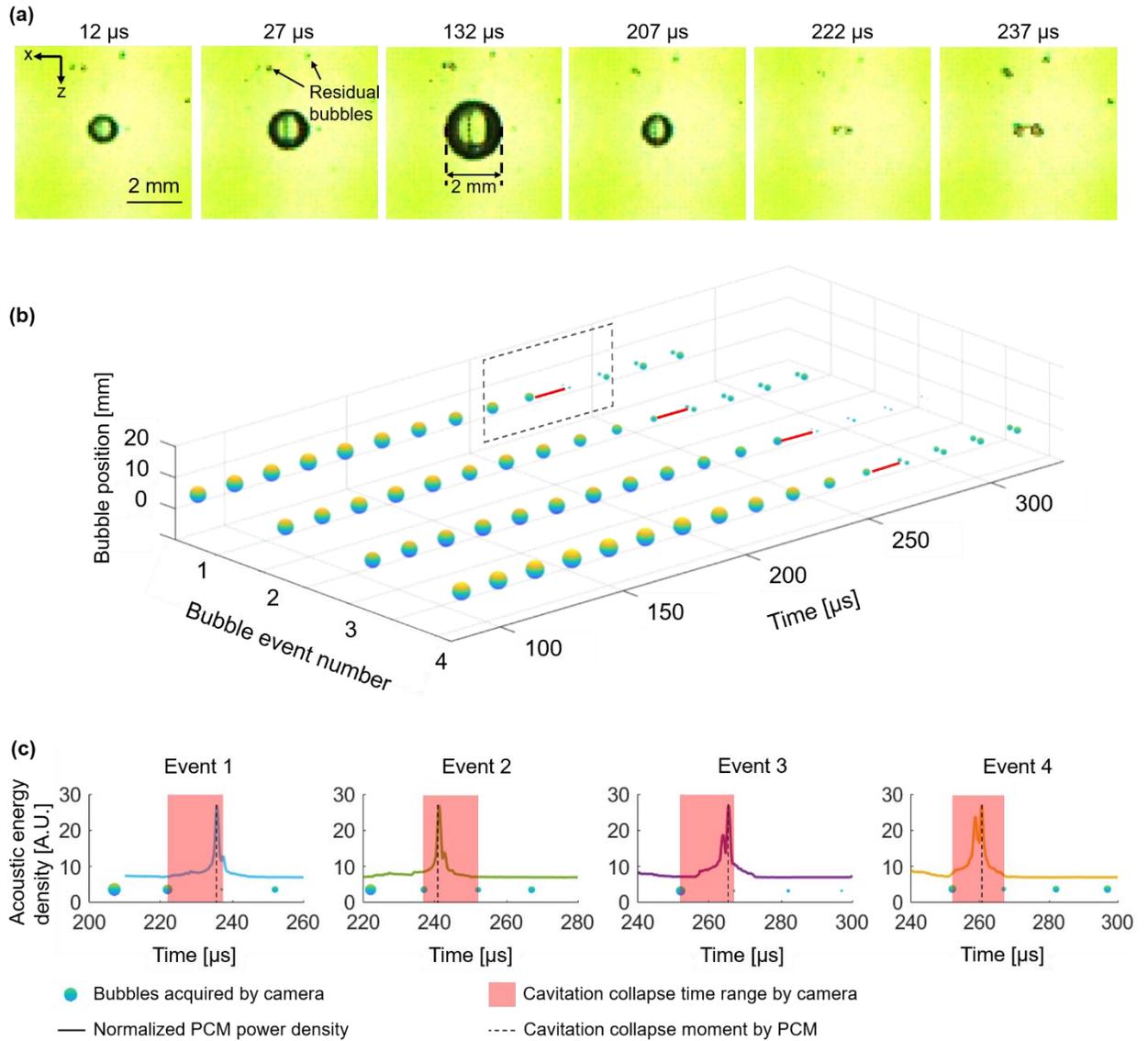

**Fig. 3**. Comparison of laser induced single bubbles in free water space detected by high-speed camera and PCM. (a) Camera-recorded single bubble images at different time instants. (b) Spatial-temporal rendering of four camera-recorded single bubble events. The corresponding red line indicates the time window of bubble collapse. (c) Close-ups of the dotted boxes from (b). The red shaded region indicates the possible time window of bubble collapse obtained by the camera, and the corresponding solid curve is the acoustic energy density extracted from the PCM reconstruction.

We further studied the bubble locations detected by ACM and PCM. While ACM measures the bubble location during expansion, PCM provides the bubble location close to its collapse. We compared the four cavitation bubble events in **Fig. 3**, in which the ACM results are shown in gray color and the corresponding PCM results are shown in color, as shown in **Fig. 4**(a). The ACM results show that the laser-induced bubbles occur consistently at nearly the same position after each laser pulse, *i.e.*, at the laser focal spot. ACM can also detect randomly distributed residual microbubbles caused by the collapse of previous bubble event, which typically have weaker reflection than the current bubble. Moreover, due to the floating effect, the residual microbubbles are usually above the current bubble. The bubble positions from the ACM and PCM images agree reasonably well for all of the cavitation events. **Figure 4**(b) illustrates the correlation between the two results of four bubble events, by showing the ACM/PCM locations and the PCM collapse time. While all the ACM images show the laser-generated bubbles at 20 μs after inception, the PCM images show the collapse of these bubbles at different times, ranging from 230 μs to 280 μs. We also analyzed 30 more bubble events to demonstrate the accuracy of PCM. In **Fig. 4**(c), all of the PCM-detected bubble collapse times are within the corresponding camera-detected collapse time windows, demonstrating PCM's high temporal accuracy. **Figure 4**(d) shows the ACM- and PCM-detected bubble locations. In the depth direction, PCM-detected bubble locations were consistently shallower than the ACM-detected locations by ~1.13 mm on average. Such discrepancy was due to both bubble expansion and floating along depth direction before its collapse. By contrast, the lateral bubble positions detected by ACM and PCM only have an average deviation of ~36.9 μm.

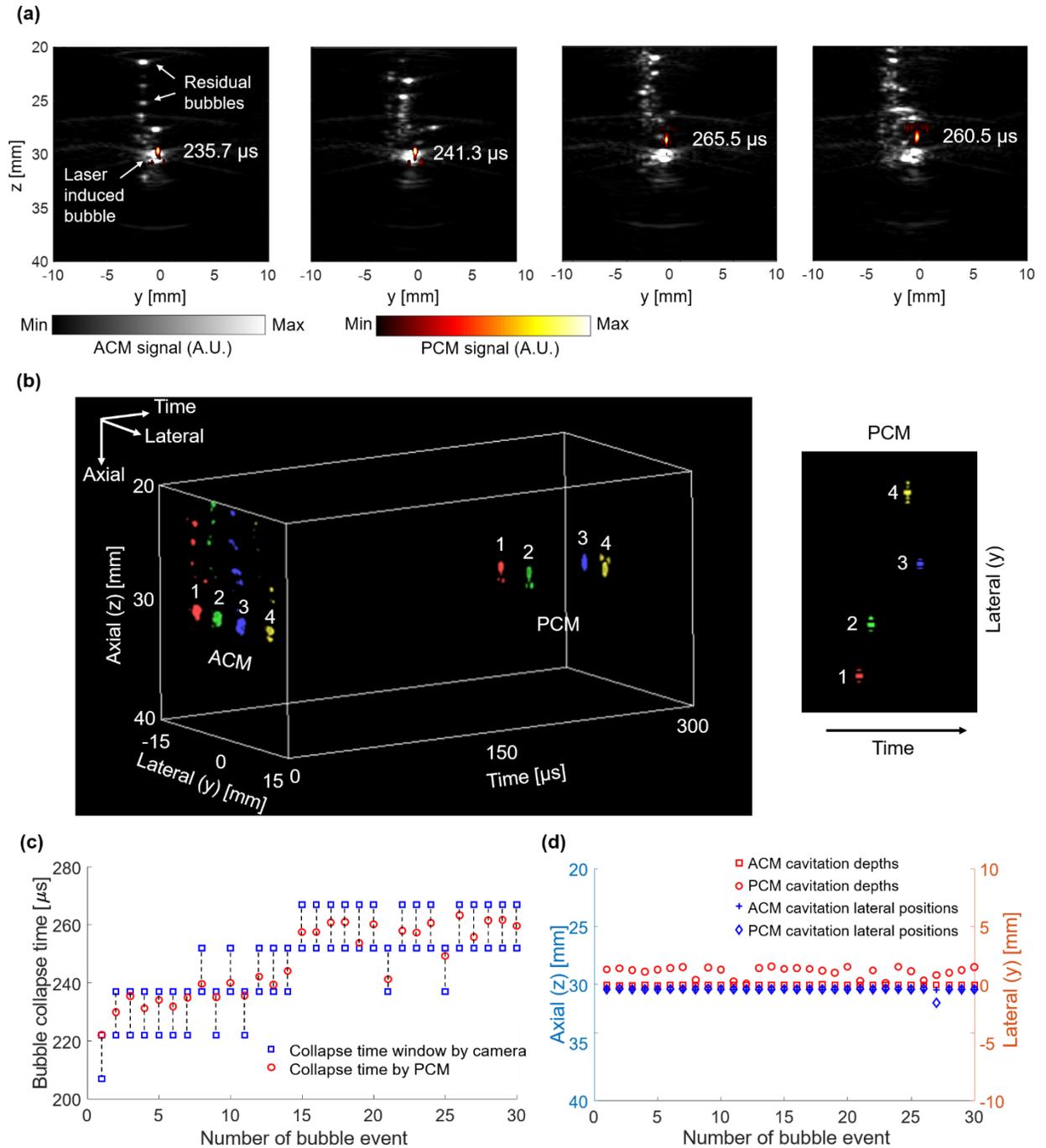

**Fig. 4**. Comparison of laser-induced single bubbles in free water space detected by ACM and PCM. (a) Superimposed ACM and PCM images of four representative bubble events. (b) Left: Spatial-temporal (lateral-axial-time) plot of bubble dynamics obtained by ACM and PCM. Right: Top view of the four PCM-detected bubbles. (c) Comparison of the bubble collapse times obtained by the camera

and PCM. (d) Comparison of bubble locations detected by ACM and PCM. The left y-axis is the axial position of the bubbles, and the right y-axis is the lateral position.

**B. Tri-modality cavitation mapping of laser-induced single bubbles in constrained space**

**Figure 5**(a) shows representative bubble dynamics in a constrained environment, obtained by the high-speed camera. Both the cavitation bubble and tube wall are clearly visible. The bubble grew to nearly the same size of the tube's inner diameter (2 mm) in 57 µs after the laser firing and collapsed between 282 µs and 297 µs. The bubble was elongated and gradually deformed into an ellipsoidal shape with an estimated major- and minor-axis of 2.3 mm and 1.7 mm, respectively, similar to the cavitation bubbles simulated in blood vessels (Farong Gao, Hu, & Hu, 2007). Subsequently, the bubble collapsed towards the right side along the tube lumen axis. In **Fig. 5**(b), we showed four representative bubble events with different collapse times. We also analyzed the acoustic energy density as a function of time from the PCM reconstruction. All PCM-detected bubble collapse times are consistent with the camera-detected time windows, as shown in **Fig. 5**(c).

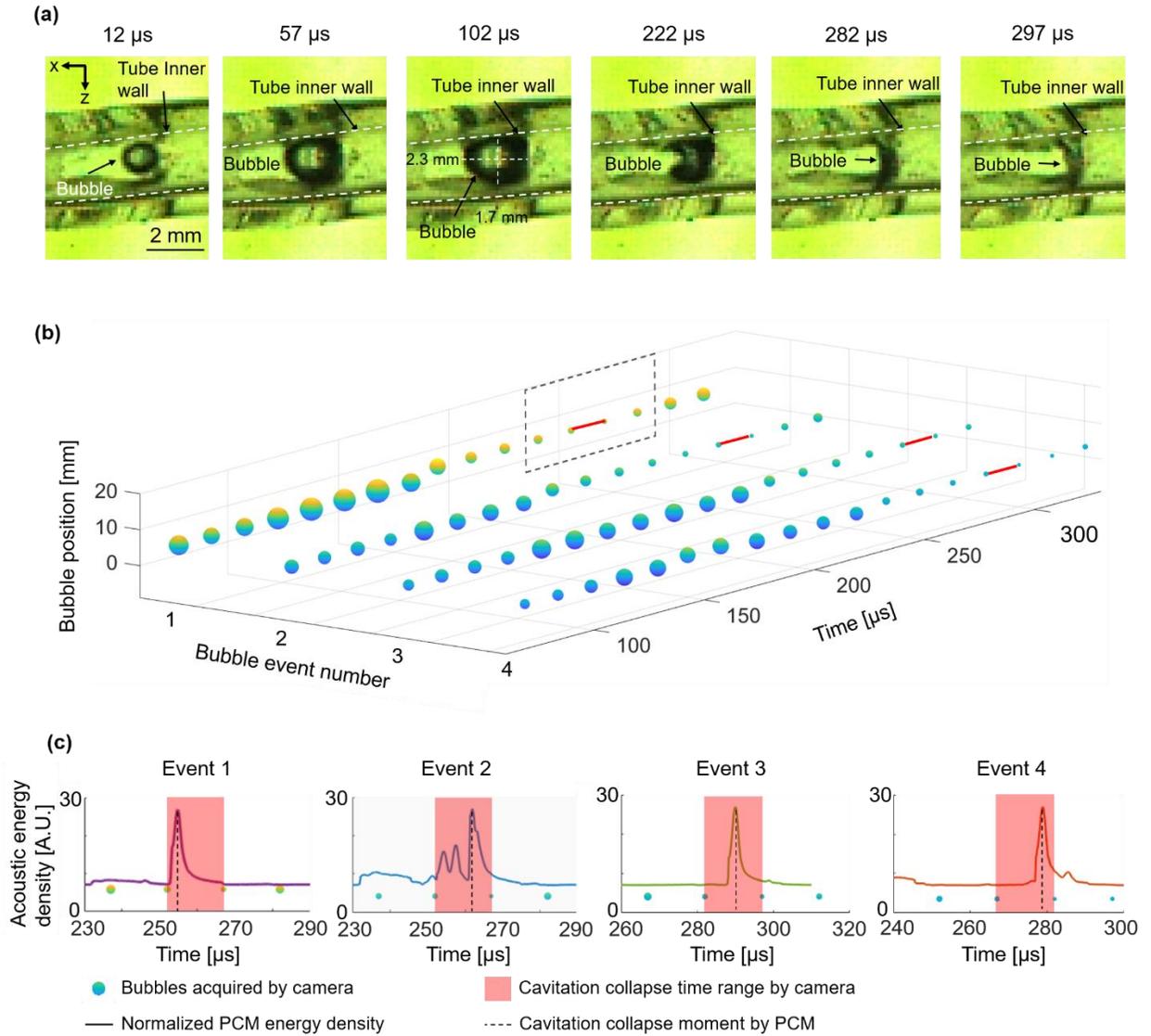

**Fig. 5**. Temporal comparison of laser induced single bubbles in a plastic tube from camera-taken photos and PCM detected results. (a) Camera-recorded single bubble images at different time instants. (b) Spatial-temporal rendering of four camera-recorded single bubble events. The corresponding red line indicates the bubble collapse time window. (c) Close-ups of the dotted boxes from (b). The red shaded region indicates the bubble collapse time window obtained by the camera, and the corresponding solid curve is the acoustic energy density extracted from the PCM reconstruction.

Again, we used ACM to validate the PCM's location accuracy, as shown in **Fig. 6**(a). The PCM detected bubble signals were weaker than that in free space, because the bubble growth in the tube is constrained and therefore less potential energy is accumulated at the maximum expansion. The asymmetric collapse also produces a weaker acoustic signal than a symmetric collapse. Additionally, the PCM results show that the bubble collapsed towards one side along the tube's lumen, which was consistent with the bubble behavior captured by the high-speed camera shown in **Fig. 5**(a). The space-time rendering plot, as shown in **Fig. 6**(b), provides the dynamics of the ACM- and PCM-detected bubbles simultaneously. We further quantified 40 bubble events in constraint space. Among the 40 bubbles, 37 PCM-detected bubble collapse times were consistent with the camera-captured time windows (**Fig. 6**(c)), resulting in a temporal accuracy of 92.5%. The depths of PCM-detected bubbles were compared with that detected by ACM, showing an average depth difference of 627 µm (**Fig. 6**(d)). The depth discrepancy was likely due to the asymmetric bubble collapse in constraint space. The lateral positions were also compared between the ACM- and PCM-detected bubbles, showing an average difference of 1.5 mm. The lateral discrepancies were larger than that in the free space due to the asymmetric bubble collapse and potentially bubble-induced streaming effect (Zhong, Cioanta, Zhu, Cocks, & Preminger, 1998).

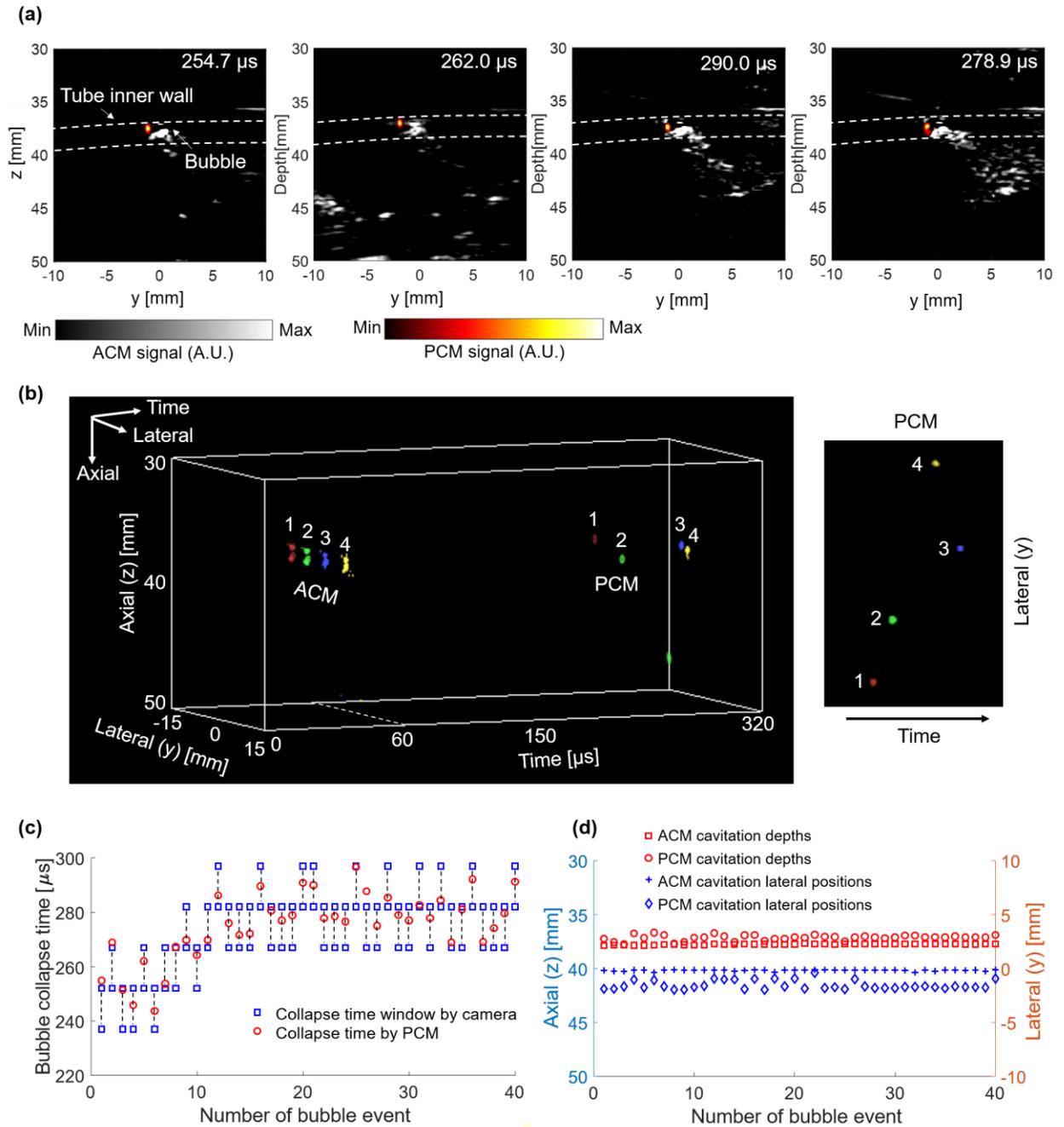

**Fig. 6**. Laser induced single bubbles in a plastic tube detected by ACM and PCM. (a) Superimposed ACM and PCM images of four representative bubble events. (b) Left: Spatial-temporal (lateral-axial-time) plot of bubble dynamics obtained by ACM and PCM. Right: Top view of the four PCM-detected bubbles. (c) Comparison of the bubble collapse times obtained by the camera and PCM. (d) Comparison of bubble locations obtained by ACM and PCM. The left y-axis is the axial position, and the right y-axis is the lateral position.

### C. Tri-modality cavitation mapping of shockwave-induced bubble clusters

**Figure 7**(a) shows a raw RGB image of the shockwave-induced bubble cluster, obtained by the high-speed camera. The red and blue channel images were extracted to provide the x-z and y-z projections of the bubble cluster, respectively. The 3D bubble distribution was then reconstructed as shown in **Fig. 7**(b), with the eleven individual bubbles clearly resolved. The simulated detection sensitivity of the ultrasound transducer in the axial direction shows a cigar-shaped detection zone across the shock-wave focal zone (**Fig. 7**(c)), in which the white-solid lines indicate the field of view (FOV) of the high-speed camera in the axial direction. We applied the acoustic detection sensitivity field to the 3D bubble distribution obtained from the camera and removed three bubbles outside of the acoustic detection zone, as shown in **Fig. 7**(d). **Figure 7**(e) shows the dynamics of the remaining eight bubbles from 120 μs to 240 μs after shockwave transmission. All the eight bubbles collapsed between 200 μs and 240 μs.

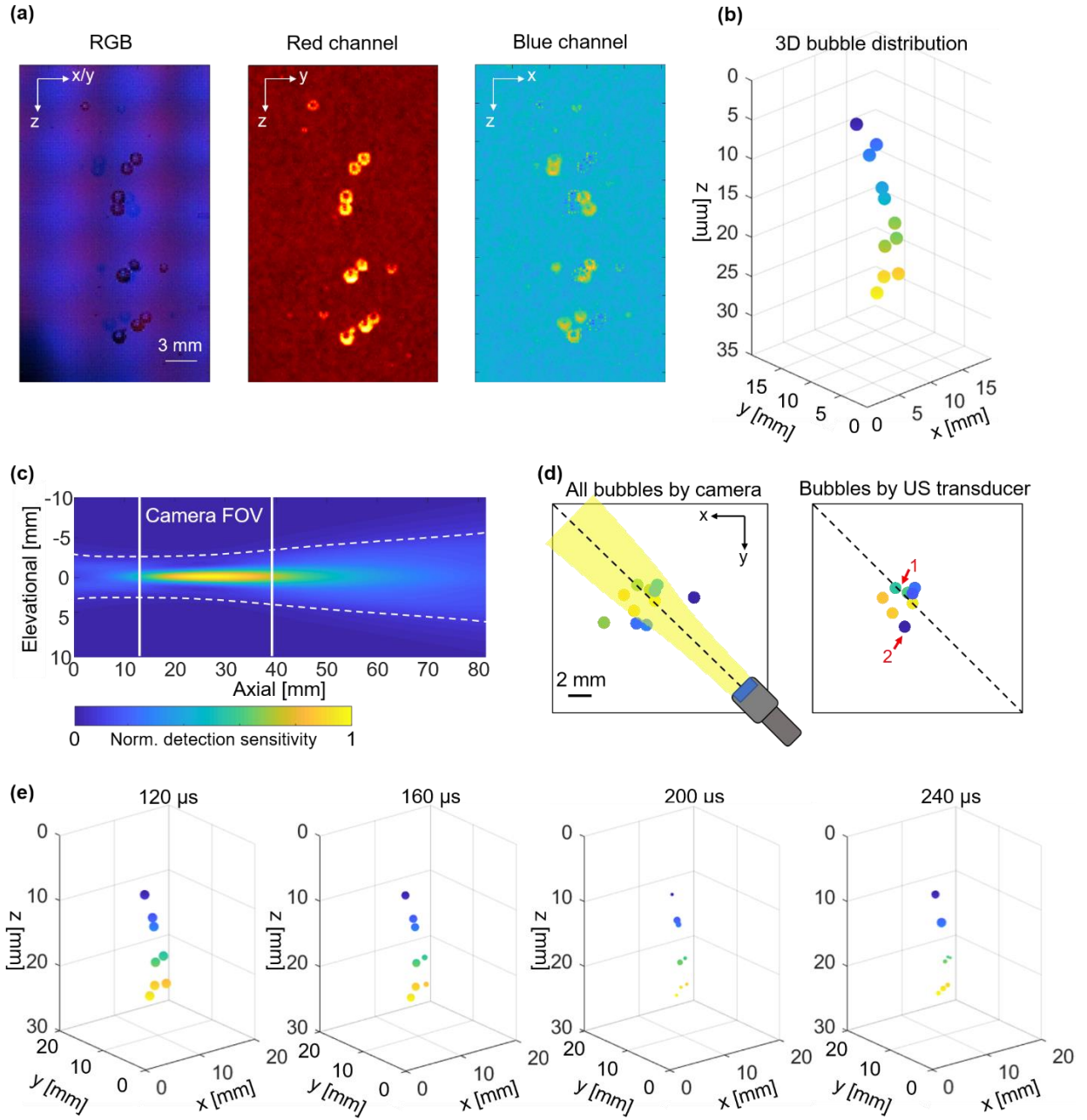

**Fig. 7**. 3D cavitation mapping of shockwave-induced bubble cluster by a high-speed camera. (a) A raw RGB image of the bubble cluster (left) and the extracted red channel image (middle) and blue channel image (right). (b) Reconstructed 3D distribution of eleven bubbles. (c) Simulated detection sensitivity of the ultrasonic transducer in the axial-elevational plane. (d) Top projections of the bubbles

before (left) and after (right) applying the ultrasound detection sensitivity. (e) 3D bubble dynamics in the acoustic detection zone.

**Figure 8**(a) shows the eight camera-captured bubbles within the ultrasonic transducer detection zone, projected onto the imaging plane of the transducer. The ACM image clearly captures the eight bubbles at the same locations as shown in the reconstructed 3D camera images. Note that the bubbles have different ACM signal amplitudes, due to the non-uniform detection sensitivity of the ultrasonic transducer. For instance, the bubble #1 was closer to the central imaging plane than the bubble #2, and thus had stronger ACM signals. We also reconstructed the bubble collapse locations from the PCM measurement (**Fig. 8**(a), right). The PCM-detected bubble collapse times were projected along the axial direction, as shown in **Fig. 8**(b), in which each bright spot represents one bubble collapse event. Although there were only eight shockwave-induced initial bubbles in the ultrasound detection zone, there were a total of 19 bubble collapse events detected by PCM, due to the second collapse of rebound bubbles. We first compared the bubble locations obtained by ACM and PCM with the high-speed camera result (**Fig. 8**(c)). The ACM-detected bubbles are highly consistent with the camera result, except for one bubble (marked by the red arrow) that may be blocked by a neighboring bubble. In the PCM image, six out of ten initial bubbles were detected by PCM and the undetected two bubbles are indicated by the green arrows. The PCM-detected bubbles do not individually match the camera captured bubbles, but they show similar group distribution. There are 4 groups of bubbles observed from the camera images. Within each group, the average diameter of bubbles is ~1 mm and the average distance between two adjacent bubbles is ~0.5 mm. Strong bubble-bubble interactions due to the secondary Bjerknes forces are therefore expected within each group of bubbles (Cheng, Hua, & Lou, 2010; Cleveland et al., 2000; Pishchalnikov, Williams, & McAteer, 2011), resulting in the displacement of the bubbles when collapsing, and thus different PCM bubble distribution. This deviation can also be observed in the camera-captured bubble collapsing. The space-time rendering

plot can simultaneously show the time and location of the bubbles detected by ACM and PCM, as shown in **Fig. 8**(d). Additional examples of cavitation analysis for shockwave induced bubble clusters are shown in **Supplementary Fig. 2**.

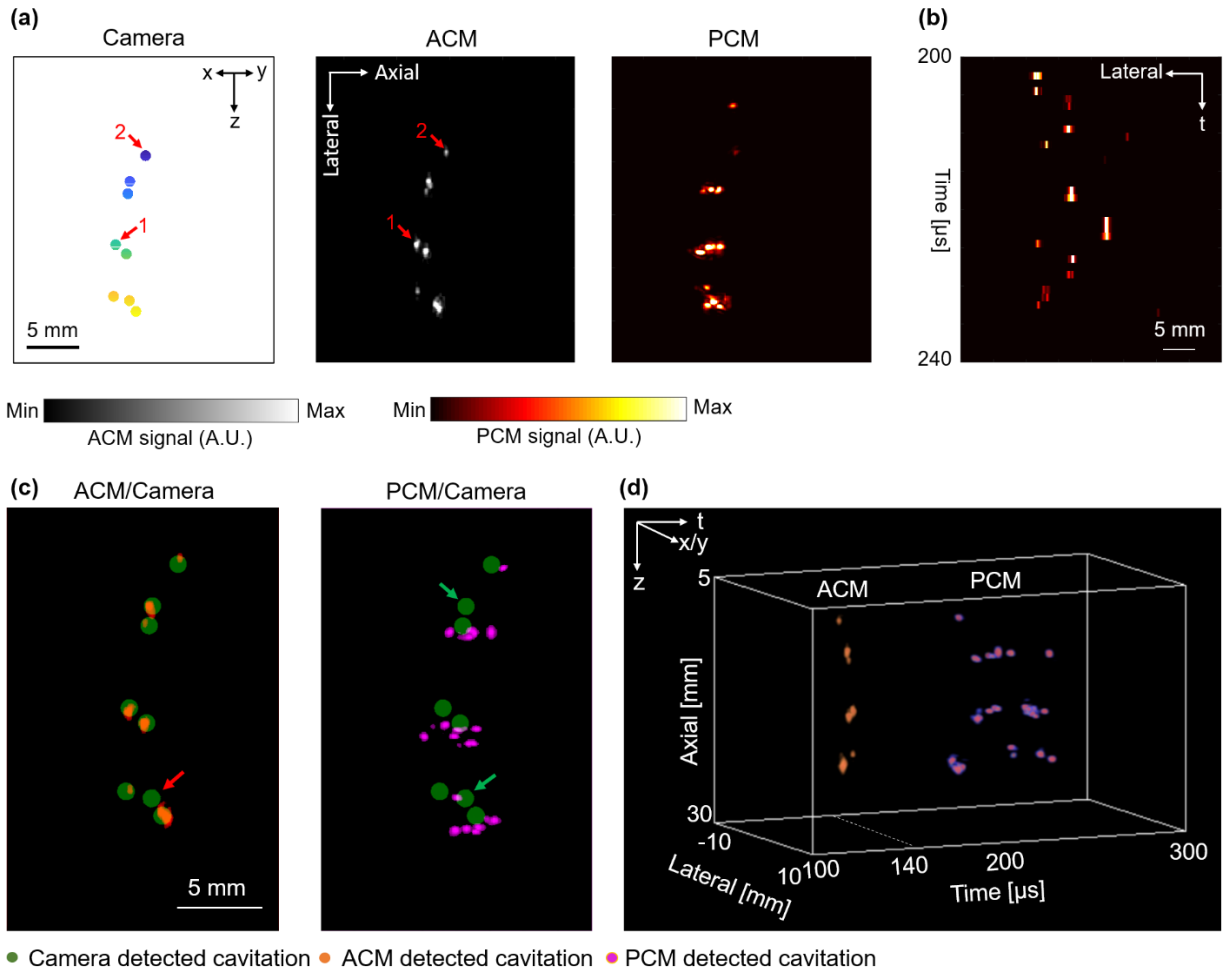

**Fig. 8**. ACM and PCM of shockwave-induced bubble clusters. (a) Cavitation bubble locations within the ultrasound detection zone, detected by high-speed camera (left), ACM (middle), and PCM (right). (b) PCM-detected bubbles projected on the lateral-time plane. (c) Superimposed images of camera-detected bubbles with the ACM- (left) and PCM-detected (right) bubbles. (d) Space-time (lateral-axial-time) rendering plot of bubble dynamics detected by ACM and PCM. The bubble distribution detected by ACM at 140 µs is shown in orange, and the bubble collapse detected by PCM is shown in magenta.

We further analyzed the bubble clusters induced by five shockwave pulses, as shown in **Table I**. For example, the first shockwave pulse generated a total of 11 bubbles. Eight bubbles were within the ultrasound detection zone, seven of which were detected by ACM. A total of 19 resultant collapse events were detected by PCM. Among them, six collapse events were from the initial bubbles and the remaining 12 collapses were from the rebound bubbles. Over the five shockwave pulses, the average bubble detection rates of ACM and PCM are 86.5% and 89.2%, respectively.

**TABLE I**. Camera/ACM/PCM detected cavitation event numbers

| Shockwave event | Total bubbles by camera | Bubbles within US detection | ACM bubbles | PCM bubbles |
| --- | --- | --- | --- | --- |
| 1 | 11 | 8 | 7 | 6 |
| 2 | 14 | 7 | 6 | 5 |
| 3 | 14 | 9 | 8 | 9 |
| 4 | 12 | 6 | 5 | 6 |
| 5 | 11 | 7 | 6 | 7 |

IV. **CONCLUSION**

Cavitation is a significant concern for tissue injury in SWL, and thus it is important to monitor the spatial-temporal distribution of cavitation bubbles to improve the therapeutic efficiency and safety. In this study, we have developed a tri-modality cavitation mapping system that seamlessly incorporates 3D high-speed camera imaging, ACM and PCM. We have thoroughly validated the PCM results by using the high-speed camera and ACM results. Our experiments of both laser- and shockwave-induced cavitation have demonstrated the high temporal and spatial accuracy of the PCM method, which is the most useful for SWL *in vivo* (Leighton et al., 2008; Zhong, Cioanta, Cocks, & Preminger, 1997). With the acquired cavitation information, urologists may be able to assess the hemorrhage potential, and thus adjust the treatment procedure and parameters to avoid severe tissue injury. It is worth noting that one drawback in this study is that the PCM uses a linear-array ultrasound probe and can only detect cavitation bubbles in the 2D imaging plane. In clinical settings, it is highly desirable to capture the shockwave-induced cavitation bubble distribution in 3D. One possible solution is to mechanically rotate the ultrasonic probe around its central axis, and thus the cavitation bubbles generated in the entire shockwave focal zone can be reconstructed at the cost of the imaging speed (Paltauf, Hartmair, Kovachev, & Nuster, 2017; Xia et al., 2013). Another solution is to apply a 2D ultrasonic probe (planar or semi-spherical) that can provide 3D cavitation mapping without mechanically scanning, at the cost of the system complexity (Dean-Ben, Fehm, Ford, Gottschalk, & Razansky, 2017; Kruger et al., 2013). Overall, future work is warranted to refine our cavitation detection methods for real-time monitoring of bubble dynamics during clinical SWL and correlation with stone fragmentation and tissue injury.

## ACKNOWLEGEMENTS

This work was supported in part by National Institute of Health (R01 EB028143, R01 NS111039, R01 NS115581, R21 EB027304, R43 CA243822, R43 CA239830, R44 HL138185, R37-DK052985-23 and P20-DK123970-01); Duke MEDx Basic Science Grant; Duke Center for Genomic and Computational Biology Faculty Research Grant; Duke Institute of Brain Science Incubator Award; and American Heart Association Collaborative Sciences Award (18CSA34080277). The authors thank Dr. Caroline Connor for editing the manuscript.